\documentclass[conference]{IEEEtran}
\IEEEoverridecommandlockouts

\usepackage{cite}
\usepackage{amsmath,amssymb,amsfonts}
\usepackage{algorithmic}
\usepackage{graphicx}
\usepackage{textcomp}
\usepackage{url}
\usepackage{booktabs}
\usepackage{xcolor}
\usepackage{subcaption}
\usepackage{multirow}

\def\BibTeX{{\rm B\kern-.05em{\sc i\kern-.025em b}\kern-.08em
    T\kern-.1667em\lower.7ex\hbox{E}\kern-.125emX}}
\begin{document}

\title{MusicEval: A Generative Music Dataset with Expert Ratings for Automatic Text-to-Music Evaluation\\
\thanks{$^{\dag}$ These authors contributed equally to this work.} \thanks{$^{*}$ Corresponding author. This work has been supported by the National Key R\&D Program of China (Grant No.2022ZD0116307) and NSF China (Grant No.62271270).}
}

\author{\IEEEauthorblockN{Cheng Liu$^{1,\dag}$, Hui Wang$^{1,\dag}$, Jinghua Zhao$^1$, Shiwan Zhao$^1$, Hui Bu$^2$, Xin Xu$^2$\\Jiaming Zhou$^1$, Haoqin Sun$^1$, and Yong Qin$^{1,*}$}
\IEEEauthorblockA{$^1$  TMCC, College of Computer Science, Nankai University, Tianjin, China, $^2$  Beijing AISHELL Technology Co., Ltd. \\
Email: liucheng\_hlt@mail.nankai.edu.cn, wanghui\_hlt@mail.nankai.edu.cn}
}

\maketitle

\begin{abstract}

The technology for generating music from textual descriptions has seen rapid advancements. However, evaluating text-to-music (TTM) systems remains a significant challenge, primarily due to the difficulty of balancing performance and cost with existing objective and subjective evaluation methods. In this paper, we propose an automatic assessment task for TTM models to align with human perception. To address the TTM evaluation challenges posed by the professional requirements of music evaluation and the complexity of the relationship between text and music, we collect MusicEval, the first generative music assessment dataset. This dataset contains 2,748 music clips generated by 31 advanced and widely used models in response to 384 text prompts, along with 13,740 ratings from 14 music experts. Furthermore, we design a CLAP-based assessment model built on this dataset, and our experimental results validate the feasibility of the proposed task,  providing a valuable reference for future development in TTM evaluation. The dataset is available at \url{https://www.aishelltech.com/AISHELL_7A}.

\end{abstract}

\begin{IEEEkeywords}
mean opinion score, text-to-music generation, automatic quality assessment
\end{IEEEkeywords}

\section{Introduction}

In recent years, music-generative models have achieved significant advances and shown considerable potential for applications in areas such as gaming and education. Among these, text-to-music (TTM) systems \cite{musiclm, musicgen, musicldm}, which generate music from natural language prompts, offer superior expression, personalization, diversity, and user-friendliness, making them more advantageous than traditional music generation systems. Many studies have utilized techniques such as diffusion models \cite{ddpm} and language models \cite{attention} to address TTM task, achieving notable results \cite{musicgen,musicldm}. However, despite the extensive focus on generative techniques, research on evaluation methods for TTM systems remains relatively underexplored. Assessing the quality of generated music is both essential and challenging as it shapes the future development of these models. Such evaluations not only play a key role in refining current-generation techniques but also stimulate innovation in model architectures and training strategies, ensuring that advancements align with the intended musical outcomes.

A universally accepted evaluation paradigm for assessing the quality of AI-generated music has not yet been established. Traditionally, methods for evaluating the quality of generative music involve a combination of objective metrics and subjective assessments. Objective metrics, such as Fréchet Audio Distance (FAD) \cite{fad} and Inception Score (IS) \cite{is}, offer quick and convenient evaluations, yet they often show a weak correlation with human-perceived quality. In contrast, subjective evaluations directly reflect human judgment, but are time-consuming, labour-intensive, and lack reproducibility, making it difficult to compare results across different works \cite{xiong,lerch22evaluating}. Therefore, there is an urgent need for an efficient and reliable method to accurately assess TTM systems.

In light of this background, we propose an automatic music quality assessment task for TTM models that aligns with human perception. Although several studies have explored automatic prediction of Mean Opinion Scores (MOS) for enhanced speech \cite{reddy2021dnsmos, mittag21_interspeech, chen2022impairment}, synthesized speech \cite{cooper2022generalization, wang23r_interspeech, wang24s_interspeech}, and voice singing \cite{singmos}, which yield promising results, the automatic evaluation of generative music aligned with human perception remains largely unexplored. Compared to the evaluation of speech quality and singing quality, the assessment of TTM systems presents unique challenges. On the one hand, assessing the quality of music involves various complex factors, such as melody, harmony, and rhythm, making it a demanding task and often requiring the opinions of professionals with specialized knowledge in music. On the other hand, compared to tasks like TTS, the relationship between the generated music and the input text descriptions in TTM systems is inherently more intricate and less direct, necessitating a deeper level of consideration and analysis. This dual challenge of evaluating both artistic quality and semantic alignment adds considerable complexity to the overall assessment process.

To address these challenges, we develop an evaluation mechanism for the TTM system based on professional music knowledge. This mechanism includes two dimensions: overall musical impression and alignment with the text prompt, which respectively emphasizes the importance of both the quality of the generated music and its consistency with the given text prompt. Based on this mechanism, we compile a generative music evaluation dataset with contributions from music experts, referred to as \textbf{MusicEval}. The MusicEval dataset contains music generated by 31 prevalent and advanced TTM models in response to 384 text prompts, with each piece of generated music scored by five experts in a back-to-back assessment. To the best of our knowledge, MusicEval is the first dataset specifically designed for the automatic evaluation of TTM systems, which offers a diverse, professional, and comprehensive data foundation for this task. Furthermore, we develop an automatic scoring system for the generated music using CLAP \cite{clap}, which assesses both the overall quality of the music and its alignment with the corresponding text prompts.

Our contributions are as follows: 
\begin{enumerate}
    \item We introduce the task of automatic evaluation for TTM systems and propose a comprehensive evaluation framework specifically tailored for TTM tasks.
    \item With the participation of music experts, we collect the first generative music assessment dataset, named \textbf{MusicEval}, which contains music from diverse and advanced systems in response to a combination of dataset-extracted and manually crafted prompts.
    \item  We develop a CLAP-based model to evaluate both the quality of generated music and its alignment with textual prompts. The experimental results demonstrate the feasibility of this task and establish a valuable baseline for future research.
\end{enumerate}

\section{The Dataset Overview}

\subsection{Basic Information}

The MusicEval dataset is a generative music evaluation dataset with a total duration of 16.62 hours, comprising 2,748 music clips in mono audio and 13,740 ratings from 14 music experts. These clips are generated by 21 different systems (spanning 31 models) in response to 384 prompts. 
To ensure consistency, we use ffmpeg to resample all generated music to a 16 kHz mono format. Each music clip is evaluated by five musical experts from conservatories, who score the clips based on two criteria: overall musical impression and alignment with the text prompts. The selection of the system and the method for designing text prompts are discussed in detail in Section~\ref{sec:Music Collection}, while the specific details of the collection and processing of scoring data are introduced in Section~\ref{sec:Mean Opinion Scores Collection}.

\subsection{Data Distribution}

\begin{figure}[t]
  \centering
  \includegraphics[scale=0.44]{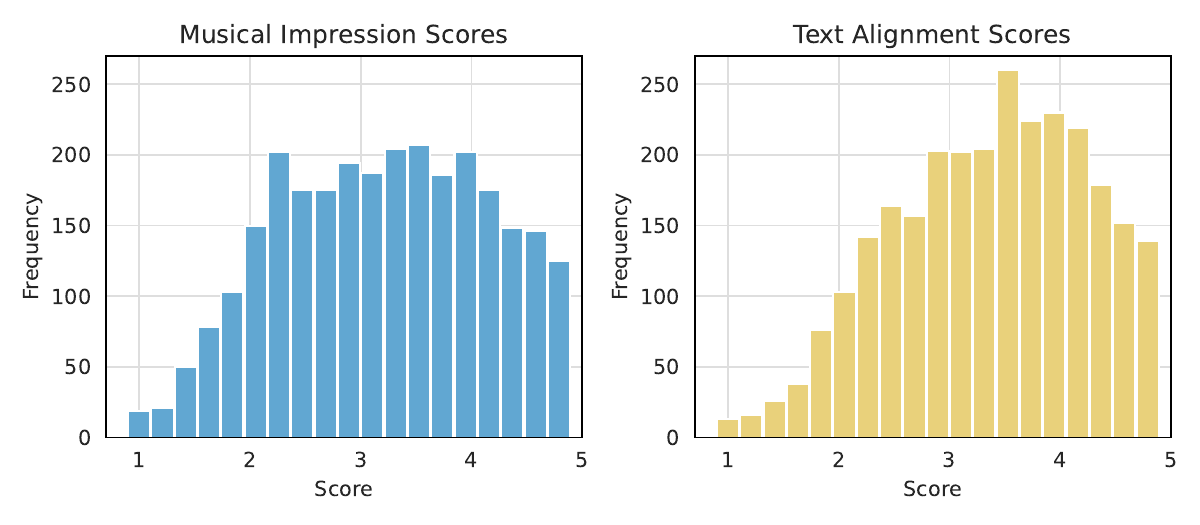}
  \caption{The distribution of musical impression scores and textual alignment scores in the MusicEval dataset.}
  \label{fig:data}
\end{figure}

Figure~\ref{fig:data} presents the distribution of overall musical impression scores and textual alignment scores, with the x-axis representing score ranges from 1 to 5 and the y-axis indicating frequency. Both distributions exhibit similar characteristics, following a roughly normal distribution pattern, with the highest frequency occurring in the mid-range scores and relatively fewer instances of extremely high or low scores. 

\begin{figure}[t]
  \centering
  \includegraphics[scale=0.55]{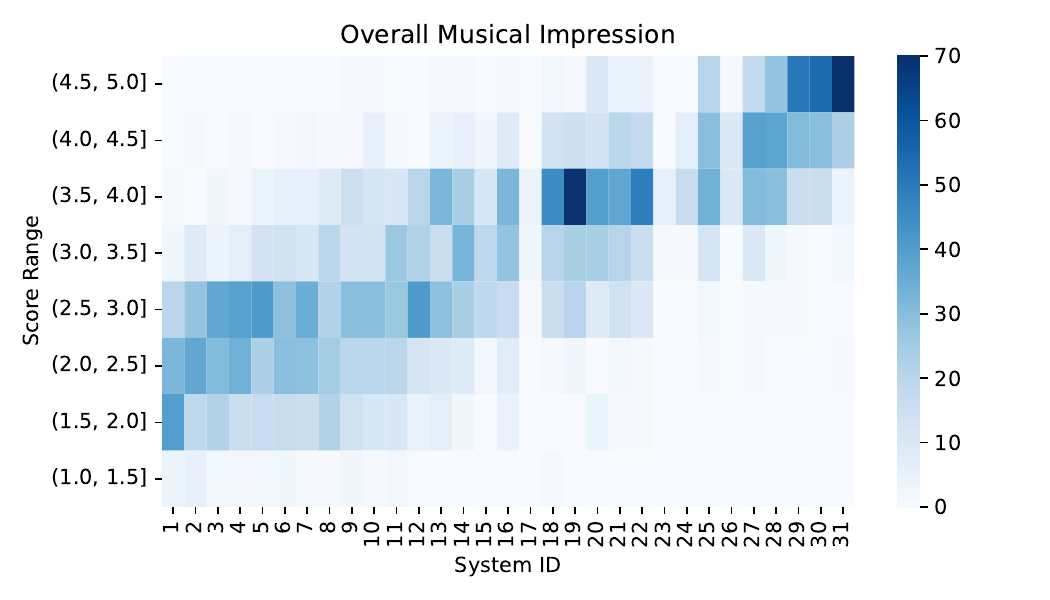}
  \caption{The distribution chart of the musical impression scores for each system in the MusicEval dataset.}
  \label{fig:qua}
\end{figure}

\begin{figure}[t]
  \centering
  \includegraphics[scale=0.55]{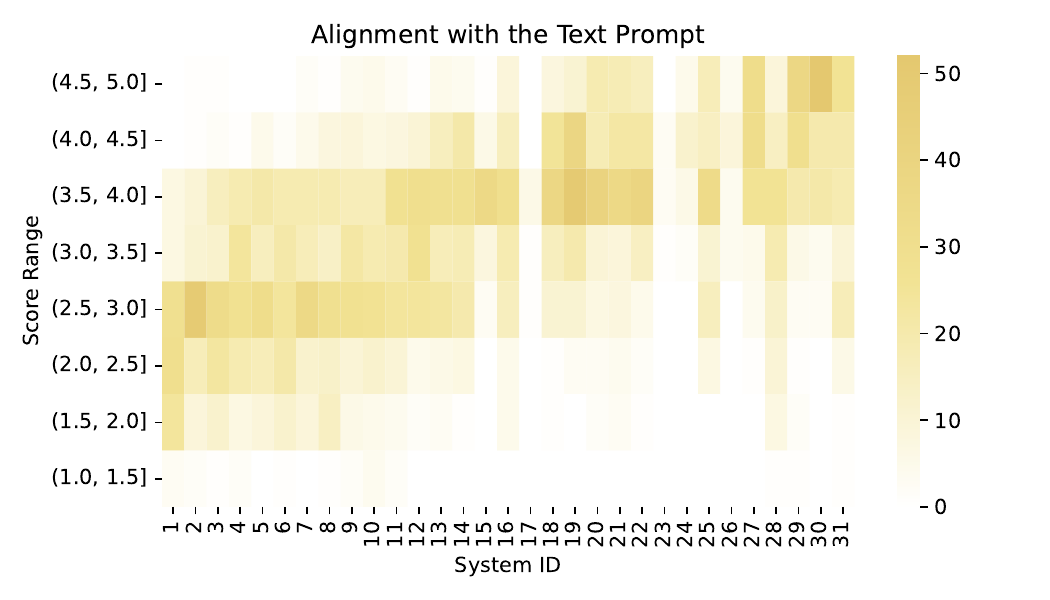}
  \caption{The distribution of the textual alignment scores for each system in the MusicEval dataset.}
  \label{fig:text}
\end{figure}

Figures~\ref{fig:qua} and~\ref{fig:text} further illustrate the distribution of scores in the two dimensions across different systems. Within each system, the generated music demonstrates relatively consistent quality; however, notable variations in quality are observed between different systems.

\section{Generated Music Collection}
\label{sec:Music Collection}

\subsection{Systems}

\begin{figure}[t]
\centering
\subfloat[The accessibility.]{
    \includegraphics[scale=0.14]{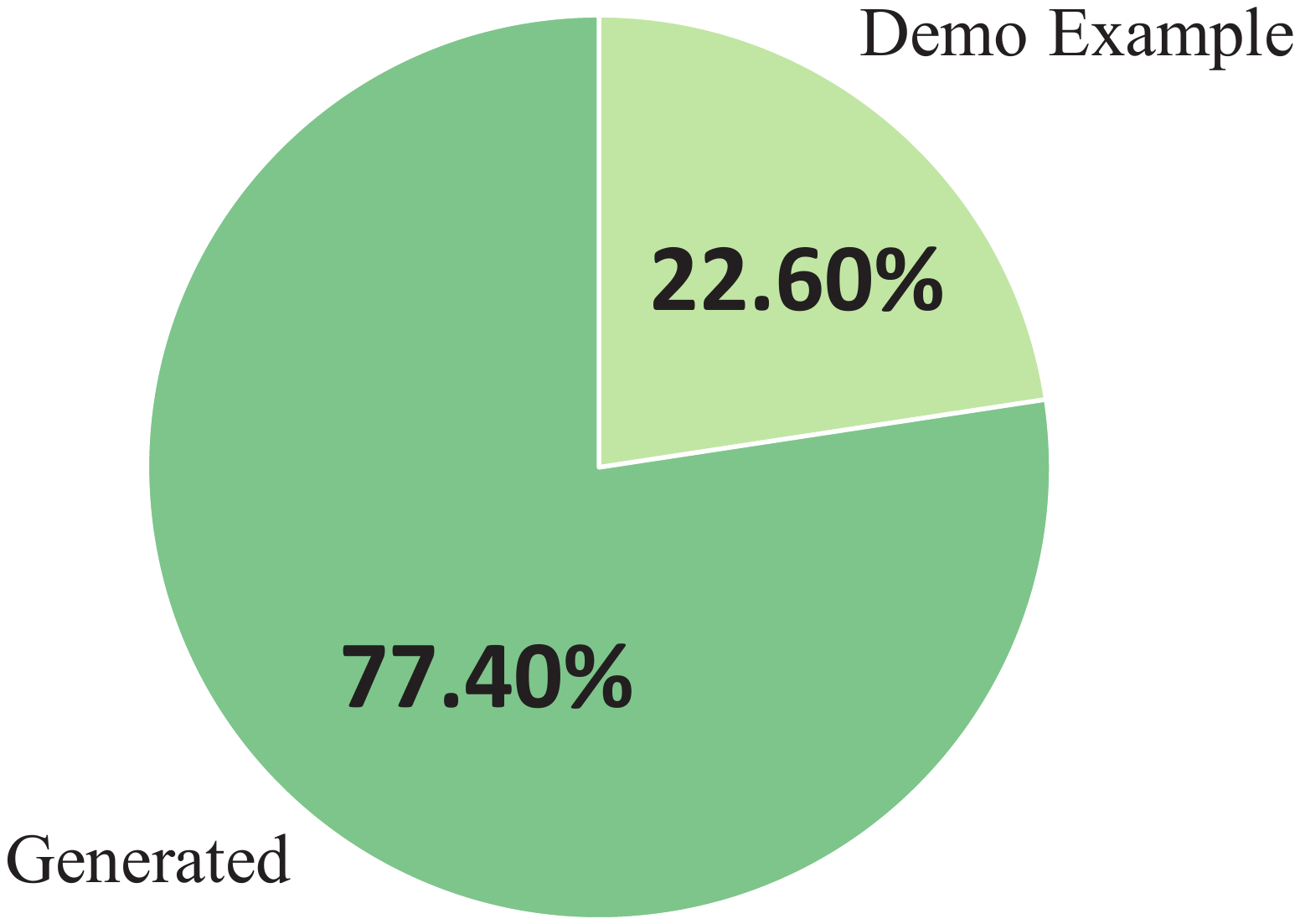}
}
\subfloat[The commercialization.]{
    \includegraphics[scale=0.14]{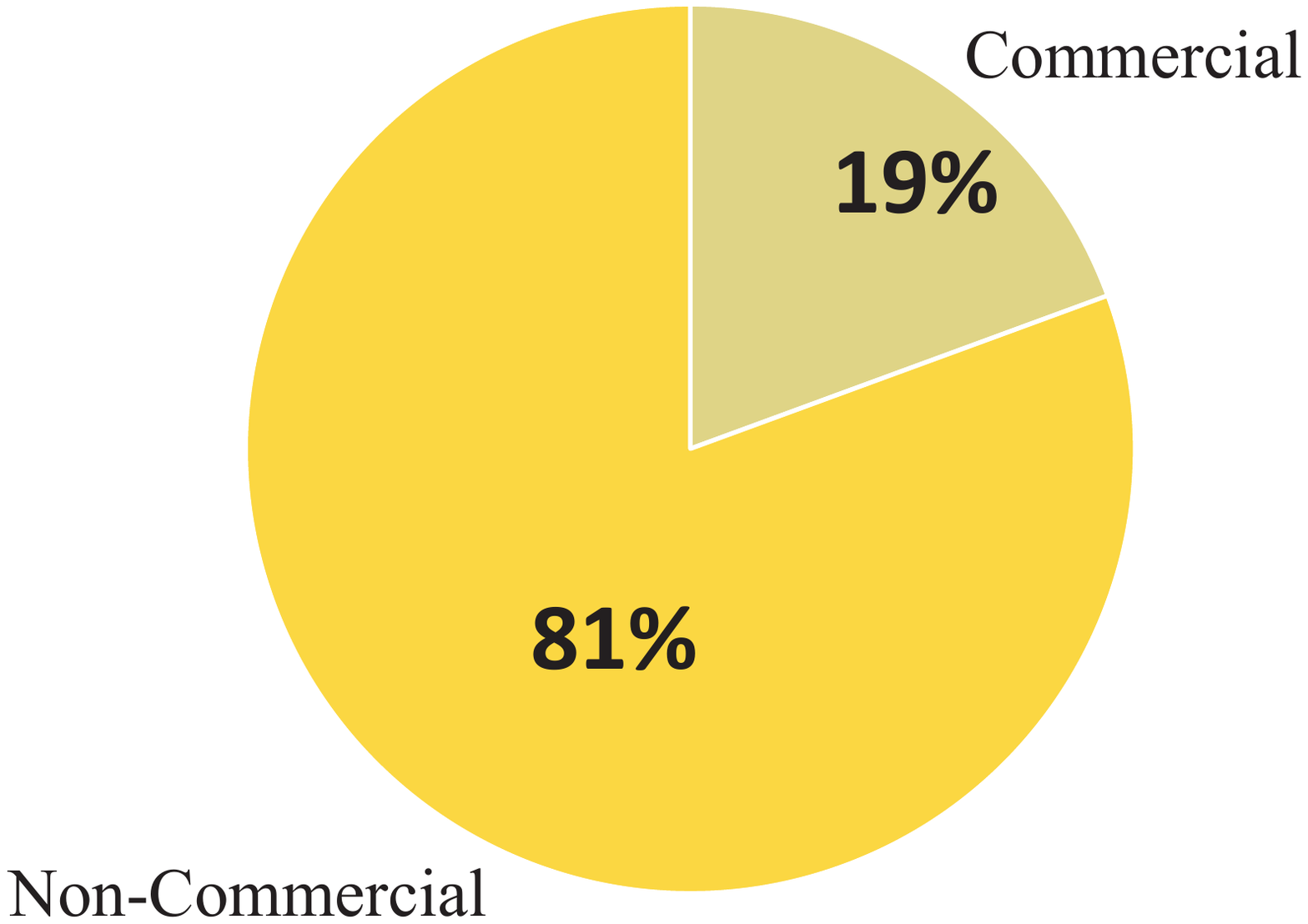}
}

\subfloat[The year.]{
    \includegraphics[scale=0.14]{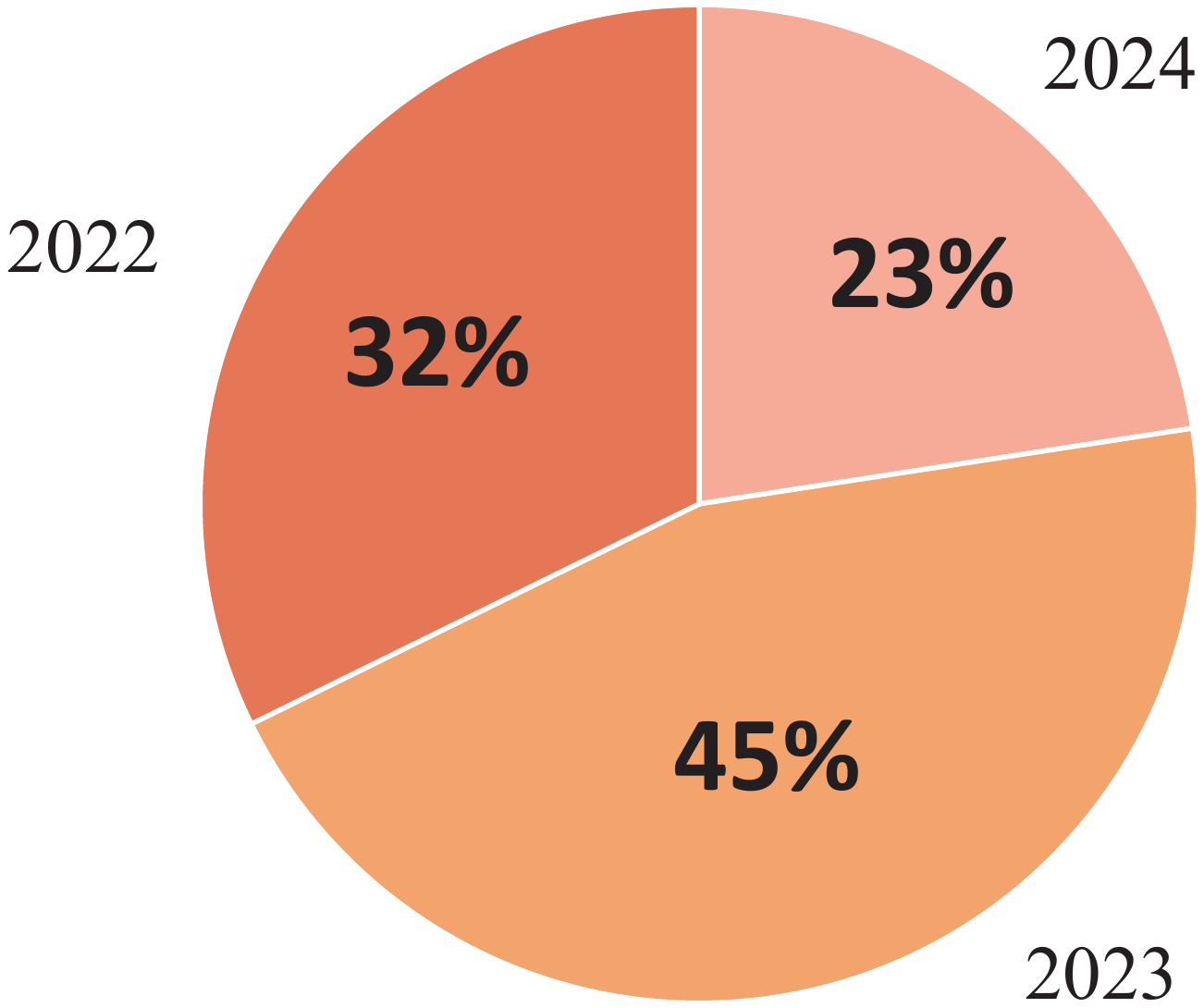}
}
\subfloat[The model size.]{
    \includegraphics[scale=0.14]{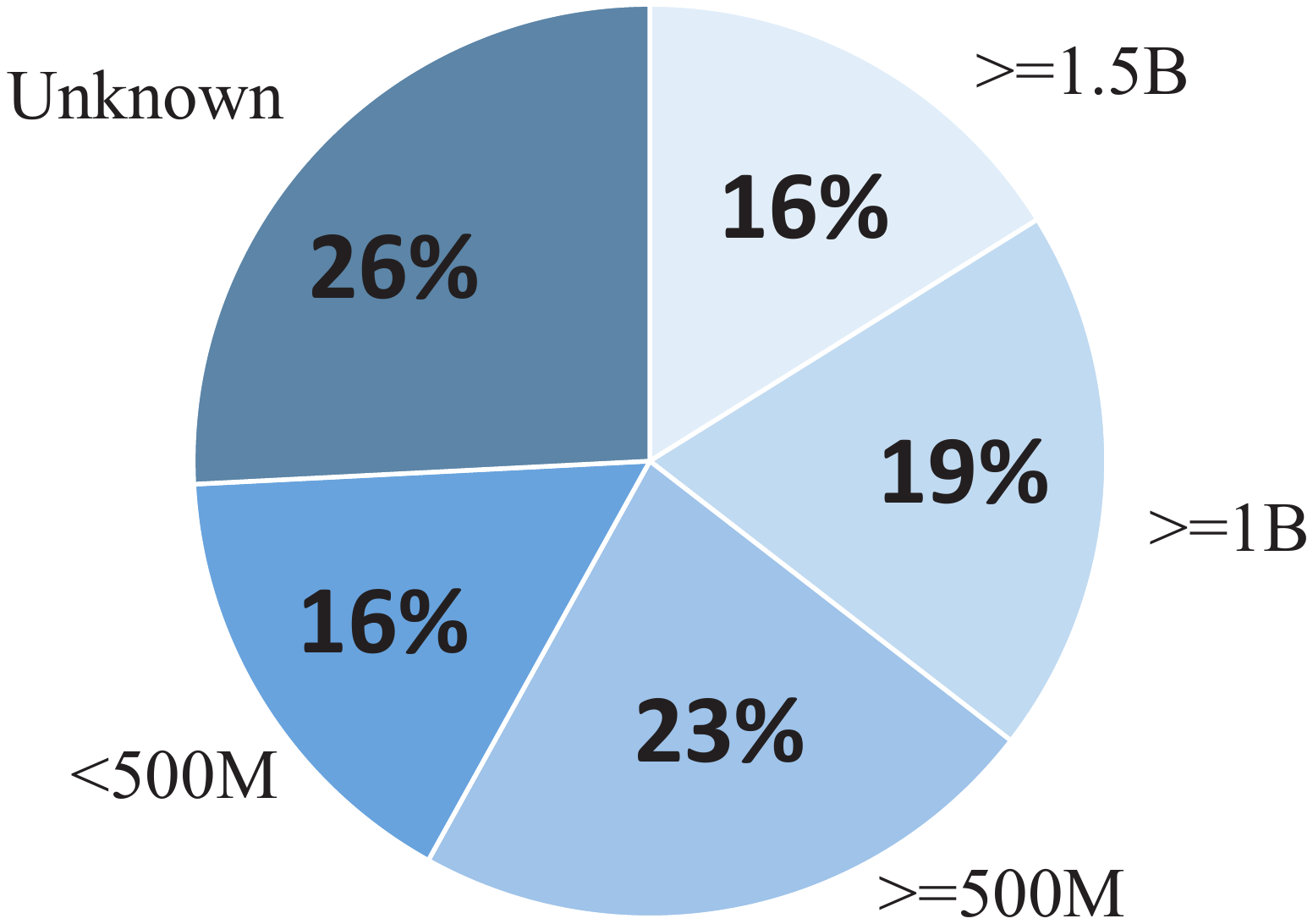}
}
\caption{The pie chart of system information across four dimensions: accessibility, commercialization, year, and model size.}
\label{fig:system}
\end{figure}
To ensure comprehensive coverage of various generative systems, we select 14 TTM systems\cite{musicgen, musicldm, musiclm, huang2023noise2music, zhu2023ernie, lu2023musecoco, magnet, mousai} and 7 text-to-audio (TTA) systems\cite{audioldm, audioldm2,tango,tango2,huang2023make,yang2023uniaudio,kreuk2022audiogen}, some of them have multiple models that differ in size or training data, resulting in a total of 31 different models. 

These systems exhibit significant variation across multiple dimensions. Figure~\ref{fig:system} presents an analysis of the 31 models across four key attributes: accessibility, commercialization, development year, and model size. In terms of accessibility, 25 systems are publicly accessible, and we used them to generate music samples based on the designed prompts. For the remaining 6 systems, which are not publicly available, we rely on their demo audio as valuable supplement data. 
Regarding commercialization, 6 systems are commercial \cite{suno,udio,sky,qqmusic,mubert,riffusion}, while 25 are non-commercial, ensuring a balanced representation of advancements from both industry and academia. The temporal distribution shows that 14 systems were developed in 2023, 7 systems were developed in 2024, and 10 in 2022, contributing to a broader temporal scope within the dataset. Additionally, the dataset includes models of varying sizes, ensuring a balanced mix of large-scale and smaller models. This diversity in multiple dimensions ensures that the dataset captures a wide range of characteristics and distributions of the TTM systems.

Most TTM systems generate music by producing latent representations or discrete tokens and decoding them into a waveform in autoregressive or non-autoregressive approaches. Besides, symbolic music generation systems have also made notable progress, although their application in TTM remains limited \cite{jisurvey} because of their inability to effectively model timbre information, as well as their heavy reliance on external components such as sound sources and rendering tools. Nevertheless, to enhance the diversity of our dataset, we include a few music samples from symbolic music generation systems \cite{lu2023musecoco} in our evaluation, including those based on ABC notation and MIDI formats.

\subsection{Text Prompts}

\begin{figure}[t]
  \centering
  \includegraphics[scale=0.44]{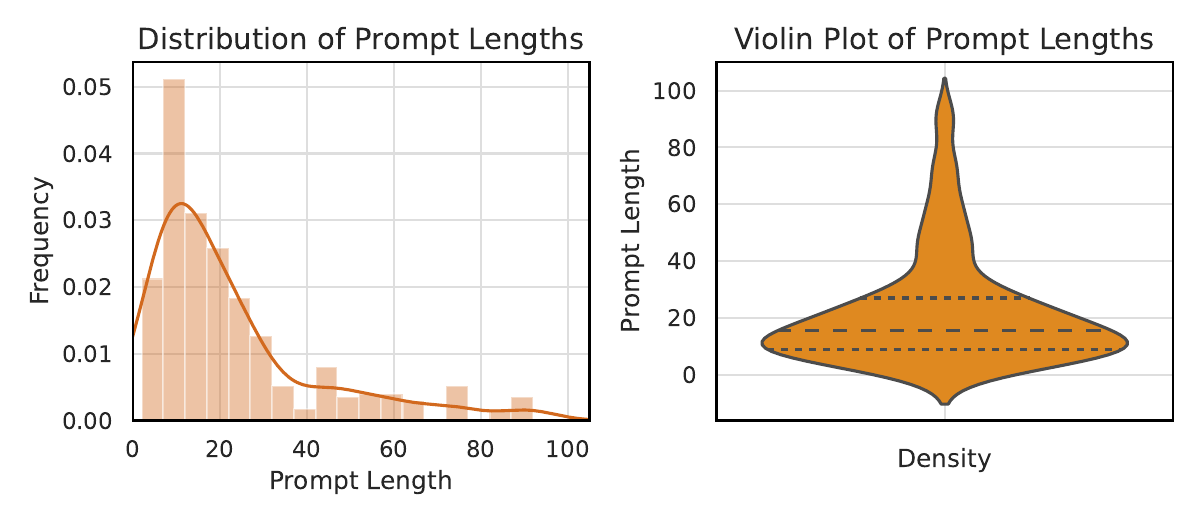}
  \caption{The distributional analysis of prompt length.}
  \label{fig:len}
\end{figure}

The TTM systems use text descriptions as input, referred to as \textit{prompts}. The MusicEval dataset includes a total of 384 prompts, comprising 80 manually crafted prompts, 20 prompts selected from the MusicCaps dataset \cite{musiclm}, and 284 prompts extracted from system demo pages. Among these, 100 text prompts are used to generate music with open-access models, while the remaining 284 prompts correspond exclusively to music clips from demo-only systems.

The manually crafted prompts are carefully designed to take multiple musical aspects into account such as emotion, structure, rhythm, theme, and instrumentation. The prompts extracted from existing datasets are included because many TTM models have encountered them during training, facilitating more likely to achieve better performance and ensuring comprehensive coverage of the MusicEval's distribution.

We analyze the distribution of prompt lengths in Figure~\ref{fig:len}, which reveals a broad range, primarily concentrated on short to mid-length prompts. The violin plot indicates high variance in prompt lengths, ensuring diversity in the generated music by accommodating both typical and atypical prompts. Moreover, the prompts in MusicEval focus on two specific genres: pop and classical music, which represent two distinct and widely recognized styles. 
Compared to other genres such as hip-hop or metal, most professional music evaluators have greater expertise in pop and classical music, enhancing the reliability of the evaluation results.

\section{Mean Opinion Scores Collection}
\label{sec:Mean Opinion Scores Collection}

We recruit 2 professional teachers and 12 experienced students from conservatories as raters and conduct a MOS test across two dimensions: \textbf{overall musical impression (musical impression)} and \textbf{alignment with the text prompt (textual alignment)}. In total, we collect 13,740 high-quality ratings from 14 raters.  

\subsection{Evaluation Dimension}

The overall musical impression score considers factors such as the authenticity of the music, and the quality of the melody, rhythm and chords. A low score indicates that the sample lacks of musicality and is of very poor quality, or exhibits noticeable machine-generated artifacts. In contrast, a high score suggests that the sample is of excellent overall quality, characterized by a clear rhythm, a pleasant melody, and a coherent chord progression, making it difficult to distinguish whether it was composed by a human or generated by a system. 

The textual alignment score assesses how well the audio sample corresponds to the given text description, reflecting the system's ability to adhere to the given prompt. A low score indicates little or no relevance between the generated music sample and the text description, while a high score suggests a strong alignment between the two.

\subsection{Listening Test Design}

All MOS tests are conducted online by specially recruited music professionals, with data packages are distributed in multiple batches. After completing a training session, the 14 rators are instructed to perform the assessments in a quiet environment. The evaluation process requires participants to read a text description and then listen to the music generated by different systems based on that description. All the music samples can be replayed repeatedly as needed. The raters are asked to score each sample on two dimensions: overall musical impression and alignment with the text prompt, and all ratings are collected using a 5-point Likert scale. Each audio sample is scored by five different evaluators, and the average score is calculated.

\subsection{Quality of Ratings}

Although subjective evaluation from human listeners is considered the gold standard, the ratings provided by music experts are not always reliable due to factors such as fatigue and distraction. To enhance the reliability of the scoring results, we adopt two probing methods in the evaluation. Firstly, several carefully selected human-created real music clips from the AudioSet\footnote{\url{https://research.google.com/audioset/index.html}} are inserted into each batch of audio data for evaluation. If a rater assigns a score lower than 3 to any of these real music samples, their scores are deemed invalid. Second, within the same batch of data, a pair of two identical audio samples are placed at different positions. If the difference in scoring results between the two samples is too large, the scores from the corresponding rater are considered invalid.

\section{Baseline}

\subsection{Model Architecture}
Inspired by the extensive success of fine-tuning pre-trained models on downstream tasks \cite{cooper2022generalization, mosssl, sun2024fine, sun24e_interspeech}, we adopt a similar approach. Specifically, we select the pre-trained CLAP model as the upstream audio feature extractor. The CLAP model is a pre-trained framework designed to learn joint embeddings of audio and text, enabling effective cross-modal understanding and retrieval between audio signals and textual descriptions \cite{laionclap2023}. We leverage 3-layer MLP as the downstream prediction head to predict musical impression scores and textual alignment scores, respectively. We guide the training process by the L1 loss between the predicted and true scores across two dimensions.

\subsection{Implementation Details}

The MusicEval data set is randomly divided into a train set and a test set in an 85\% and 15\% ratio as a preliminary split setting. For the CLAP model, we utilize the official pre-trained checkpoint\footnote{\url{https://huggingface.co/lukewys/laion_clap/blob/main/music_audioset_epoch_15_esc_90.14.pt}} which is trained on music, AudioSet, and LAION-Audio-630k. During the fine-tuning stage, we conduct the experiments using a batch size of 64 on a single NVIDIA 4090 GPU. The SGD optimizer is employed with a learning rate of 0.0005.

\subsection{Test Metrics}

We evaluate the performance by employing several objective indicators. We evaluate the Mean square error (MSE), the Linear Correlation Coefficient (LCC), the Spearman Rank Correlation Coefficient (SRCC) and the Kendall Tau Rank Correlation (KTAU)\cite{cooper2022generalization} between predictions and ground-truths.

\subsection{Experimental Results}
As shown in Table~\ref{tab:res}, we evaluate our model on the MusicEval test set across two dimensions: Musical impression and Textual alignment. For the first evaluation dimension at utterance-level and system-level, the MSE is relatively low, while the LCC, SRCC, and KTAU are all high, indicating a strong correlation with human evaluations. In contrast, the second result shows slightly higher MSE and lower correlation metrics, suggesting that there is more difficulty in predicting textual alignment compared to the musical impression. In general, the results demonstrate that the scores predicted by the fine-tuned prediction model are strongly correlated with those of human experts, confirming the feasibility of using deep neural networks to automatically assess the quality of the generated music.

\begin{table}[t]
\centering
\caption{Performance of our baseline model on the MusicEval test set. The prefix \textbf{U\_} represents utterance-level metrics, while the prefix \textbf{S\_} denotes system-level metrics.}
\label{tab:res}
\begin{tabular}{ccccc}
\toprule
& U\_MSE↓ & U\_LCC↑ & U\_SRCC↑ & U\_KATU↑ \\
\midrule
Musical impression & 0.647 & 0.606 & 0.633 & 0.461 \\
Textual alignment & 0.616 & 0.438 & 0.443 & 0.314 \\
\midrule\midrule
& S\_MSE↓ & S\_LCC↑ & S\_SRCC↑ & S\_KATU↑ \\
\midrule
Musical impression & 0.446 & 0.839 & 0.926 & 0.767 \\
Textual alignment & 0.354 & 0.757 & 0.784 & 0.617 \\
\bottomrule
\end{tabular}
\end{table}

\section{Conclusion}
In this paper, we present MusicEval, the first expert-scored dataset for generative music evaluation, which includes a wide range of music generated by diverse TTM systems and MOS ratings on a Likert scale from music experts. Furthermore, we propose a baseline model utilizing CLAP to predict both overall musical impressions and alignment with textual descriptions of generated music. Our results demonstrate the effectiveness of this automatic evaluation method. In future work, we aim to enrich the dataset to encompass more additional genres and explore more advanced architectures for the automatic evaluation of TTM.

\bibliographystyle{IEEEbib}
\bibliography{refs}

\begin{thebibliography}{10}

\bibitem{musiclm}
Andrea Agostinelli, Timo~I Denk, Zal{\'a}n Borsos, Jesse Engel, Mauro Verzetti, Antoine Caillon, Qingqing Huang, Aren Jansen, Adam Roberts, Marco Tagliasacchi, et~al.,
\newblock ``Musiclm: Generating music from text,''
\newblock {\em arXiv preprint arXiv:2301.11325}, 2023.

\bibitem{musicgen}
Jade Copet, Felix Kreuk, Itai Gat, Tal Remez, David Kant, Gabriel Synnaeve, Yossi Adi, and Alexandre Defossez,
\newblock ``Simple and controllable music generation,''
\newblock in {\em Advances in Neural Information Processing Systems}, A.~Oh, T.~Naumann, A.~Globerson, K.~Saenko, M.~Hardt, and S.~Levine, Eds. 2023, vol.~36, pp. 47704--47720, Curran Associates, Inc.

\bibitem{musicldm}
Ke~Chen, Yusong Wu, Haohe Liu, Marianna Nezhurina, Taylor Berg-Kirkpatrick, and Shlomo Dubnov,
\newblock ``Musicldm: Enhancing novelty in text-to-music generation using beat-synchronous mixup strategies,''
\newblock in {\em ICASSP 2024 - 2024 IEEE International Conference on Acoustics, Speech and Signal Processing (ICASSP)}, 2024, pp. 1206--1210.

\bibitem{ddpm}
Jonathan Ho, Ajay Jain, and Pieter Abbeel,
\newblock ``Denoising diffusion probabilistic models,''
\newblock in {\em Advances in Neural Information Processing Systems}, H.~Larochelle, M.~Ranzato, R.~Hadsell, M.F. Balcan, and H.~Lin, Eds. 2020, vol.~33, pp. 6840--6851, Curran Associates, Inc.

\bibitem{attention}
Ashish Vaswani, Noam Shazeer, Niki Parmar, Jakob Uszkoreit, Llion Jones, Aidan~N Gomez, \L~ukasz Kaiser, and Illia Polosukhin,
\newblock ``Attention is all you need,''
\newblock in {\em Advances in Neural Information Processing Systems}, I.~Guyon, U.~Von Luxburg, S.~Bengio, H.~Wallach, R.~Fergus, S.~Vishwanathan, and R.~Garnett, Eds. 2017, vol.~30, Curran Associates, Inc.

\bibitem{fad}
Kevin Kilgour, Mauricio Zuluaga, Dominik Roblek, and Matthew Sharifi,
\newblock ``Fr\'echet audio distance: A metric for evaluating music enhancement algorithms,'' 2019.

\bibitem{is}
Shane Barratt and Rishi Sharma,
\newblock ``A note on the inception score,''
\newblock {\em arXiv preprint arXiv:1801.01973}, 2018.

\bibitem{xiong}
Zeyu Xiong, Weitao Wang, Jing Yu, Yue Lin, and Ziyan Wang,
\newblock ``A comprehensive survey for evaluation methodologies of ai-generated music,'' 2023.

\bibitem{lerch22evaluating}
Ashvala Vinay and Alexander Lerch,
\newblock ``Evaluating generative audio systems and their metrics,''
\newblock {\em arXiv preprint arXiv:2209.00130}, 2022.

\bibitem{reddy2021dnsmos}
Chandan~KA Reddy, Vishak Gopal, and Ross Cutler,
\newblock ``Dnsmos: A non-intrusive perceptual objective speech quality metric to evaluate noise suppressors,''
\newblock in {\em ICASSP 2021-2021 IEEE International Conference on Acoustics, Speech and Signal Processing (ICASSP)}. IEEE, 2021, pp. 6493--6497.

\bibitem{mittag21_interspeech}
Gabriel Mittag, Babak Naderi, Assmaa Chehadi, and Sebastian Möller,
\newblock ``Nisqa: A deep cnn-self-attention model for multidimensional speech quality prediction with crowdsourced datasets,''
\newblock in {\em Interspeech 2021}, 2021, pp. 2127--2131.

\bibitem{chen2022impairment}
Lianwu Chen, Xinlei Ren, Xu~Zhang, Xiguang Zheng, Chen Zhang, Liang Guo, and Bing Yu,
\newblock ``Impairment representation learning for speech quality assessment.,''
\newblock in {\em INTERSPEECH}, 2022, pp. 3323--3327.

\bibitem{cooper2022generalization}
Erica Cooper, Wen-Chin Huang, Tomoki Toda, and Junichi Yamagishi,
\newblock ``Generalization ability of mos prediction networks,''
\newblock in {\em ICASSP 2022-2022 IEEE International Conference on Acoustics, Speech and Signal Processing (ICASSP)}. IEEE, 2022, pp. 8442--8446.

\bibitem{wang23r_interspeech}
Hui Wang, Shiwan Zhao, Xiguang Zheng, and Yong Qin,
\newblock ``Ramp: Retrieval-augmented mos prediction via confidence-based dynamic weighting,''
\newblock in {\em INTERSPEECH 2023}, 2023, pp. 1095--1099.

\bibitem{wang24s_interspeech}
Hui Wang, Shiwan Zhao, Jiaming Zhou, Xiguang Zheng, Haoqin Sun, Xuechen Wang, and Yong Qin,
\newblock ``Uncertainty-aware mean opinion score prediction,''
\newblock in {\em Interspeech 2024}, 2024, pp. 1215--1219.

\bibitem{singmos}
Yuxun Tang, Jiatong Shi, Yuning Wu, and Qin Jin,
\newblock ``Singmos: An extensive open-source singing voice dataset for mos prediction,'' 2024.

\bibitem{clap}
Yusong Wu, Ke~Chen, Tianyu Zhang, Yuchen Hui, Taylor Berg-Kirkpatrick, and Shlomo Dubnov,
\newblock ``Large-scale contrastive language-audio pretraining with feature fusion and keyword-to-caption augmentation,''
\newblock in {\em ICASSP 2023 - 2023 IEEE International Conference on Acoustics, Speech and Signal Processing (ICASSP)}, 2023, pp. 1--5.

\bibitem{huang2023noise2music}
Qingqing Huang, Daniel~S Park, Tao Wang, Timo~I Denk, Andy Ly, Nanxin Chen, Zhengdong Zhang, Zhishuai Zhang, Jiahui Yu, Christian Frank, et~al.,
\newblock ``Noise2music: Text-conditioned music generation with diffusion models,''
\newblock {\em arXiv preprint arXiv:2302.03917}, 2023.

\bibitem{zhu2023ernie}
Pengfei Zhu, Chao Pang, Yekun Chai, Lei Li, Shuohuan Wang, Yu~Sun, Hao Tian, and Hua Wu,
\newblock ``Ernie-music: Text-to-waveform music generation with diffusion models,''
\newblock {\em arXiv preprint arXiv:2302.04456}, 2023.

\bibitem{lu2023musecoco}
Peiling Lu, Xin Xu, Chenfei Kang, Botao Yu, Chengyi Xing, Xu~Tan, and Jiang Bian,
\newblock ``Musecoco: Generating symbolic music from text,''
\newblock {\em arXiv preprint arXiv:2306.00110}, 2023.

\bibitem{magnet}
Alon Ziv, Itai Gat, Gael~Le Lan, Tal Remez, Felix Kreuk, Alexandre D{\'e}fossez, Jade Copet, Gabriel Synnaeve, and Yossi Adi,
\newblock ``Masked audio generation using a single non-autoregressive transformer,''
\newblock {\em arXiv preprint arXiv:2401.04577}, 2024.

\bibitem{mousai}
Flavio Schneider, Ojasv Kamal, Zhijing Jin, and Bernhard Schölkopf,
\newblock ``Mo\^usai: Text-to-music generation with long-context latent diffusion,'' 2023.

\bibitem{audioldm}
Haohe Liu, Zehua Chen, Yiitan Yuan, Xinhao Mei, Xubo Liu, Danilo~P. Mandic, Wenwu Wang, and MarkD~. Plumbley,
\newblock ``Audioldm: Text-to-audio generation with latent diffusion models,''
\newblock in {\em International Conference on Machine Learning}, 2023.

\bibitem{audioldm2}
Haohe Liu, Yi~Yuan, Xubo Liu, Xinhao Mei, Qiuqiang Kong, Qiao Tian, Yuping Wang, Wenwu Wang, Yuxuan Wang, and Mark~D Plumbley,
\newblock ``Audioldm 2: Learning holistic audio generation with self-supervised pretraining,''
\newblock {\em IEEE/ACM Transactions on Audio, Speech, and Language Processing}, 2024.

\bibitem{tango}
Deepanway Ghosal, Navonil Majumder, Ambuj Mehrish, and Soujanya Poria,
\newblock ``Text-to-audio generation using instruction guided latent diffusion model,''
\newblock in {\em Proceedings of the 31st ACM International Conference on Multimedia}, New York, NY, USA, 2023, MM '23, p. 3590–3598, Association for Computing Machinery.

\bibitem{tango2}
Navonil Majumder, Chia-Yu Hung, Deepanway Ghosal, Wei-Ning Hsu, Rada Mihalcea, and Soujanya Poria,
\newblock ``Tango 2: Aligning diffusion-based text-to-audio generations through direct preference optimization,''
\newblock {\em arXiv preprint arXiv:2404.09956}, 2024.

\bibitem{huang2023make}
Rongjie Huang, Jiawei Huang, Dongchao Yang, Yi~Ren, Luping Liu, Mingze Li, Zhenhui Ye, Jinglin Liu, Xiang Yin, and Zhou Zhao,
\newblock ``Make-an-audio: Text-to-audio generation with prompt-enhanced diffusion models,''
\newblock in {\em International Conference on Machine Learning}. PMLR, 2023, pp. 13916--13932.

\bibitem{yang2023uniaudio}
Dongchao Yang, Jinchuan Tian, Xu~Tan, Rongjie Huang, Songxiang Liu, Xuankai Chang, Jiatong Shi, Sheng Zhao, Jiang Bian, Xixin Wu, et~al.,
\newblock ``Uniaudio: An audio foundation model toward universal audio generation,''
\newblock {\em arXiv preprint arXiv:2310.00704}, 2023.

\bibitem{kreuk2022audiogen}
Felix Kreuk, Gabriel Synnaeve, Adam Polyak, Uriel Singer, Alexandre D{\'e}fossez, Jade Copet, Devi Parikh, Yaniv Taigman, and Yossi Adi,
\newblock ``Audiogen: Textually guided audio generation,''
\newblock {\em arXiv preprint arXiv:2209.15352}, 2022.

\bibitem{suno}
Suno,
\newblock ``Suno official website,'' 2024.

\bibitem{udio}
Udio,
\newblock ``Udio official website,'' 2024.

\bibitem{sky}
SkyMusic,
\newblock ``Skymusic official website,'' 2024.

\bibitem{qqmusic}
qqmusic,
\newblock ``qqmusic official website,'' 2024.

\bibitem{mubert}
mubert,
\newblock ``mubert official website,'' 2024.

\bibitem{riffusion}
riffusion,
\newblock ``riffusion official website,'' 2024.

\bibitem{jisurvey}
Shulei Ji, Xinyu Yang, and Jing Luo,
\newblock ``A survey on deep learning for symbolic music generation: Representations, algorithms, evaluations, and challenges,''
\newblock {\em ACM Computing Surveys}, vol. 56, no. 1, pp. 1--39, 2023.

\bibitem{mosssl}
Wei-Cheng Tseng, Chien yu~Huang, Wei-Tsung Kao, Yist~Y. Lin, and Hung yi~Lee,
\newblock ``Utilizing self-supervised representations for mos prediction,'' 2021.

\bibitem{sun2024fine}
Haoqin Sun, Shiwan Zhao, Xuechen Wang, Wenjia Zeng, Yong Chen, and Yong Qin,
\newblock ``Fine-grained disentangled representation learning for multimodal emotion recognition,''
\newblock in {\em ICASSP 2024-2024 IEEE International Conference on Acoustics, Speech and Signal Processing (ICASSP)}. IEEE, 2024, pp. 11051--11055.

\bibitem{sun24e_interspeech}
Haoqin Sun, Shiwan Zhao, Xiangyu Kong, Xuechen Wang, Hui Wang, Jiaming Zhou, and Yong Qin,
\newblock ``Iterative prototype refinement for ambiguous speech emotion recognition,''
\newblock in {\em Interspeech 2024}, 2024, pp. 3200--3204.

\bibitem{laionclap2023}
Yusong Wu*, Ke~Chen*, Tianyu Zhang*, Yuchen Hui*, Taylor Berg-Kirkpatrick, and Shlomo Dubnov,
\newblock ``Large-scale contrastive language-audio pretraining with feature fusion and keyword-to-caption augmentation,''
\newblock in {\em IEEE International Conference on Acoustics, Speech and Signal Processing, ICASSP}, 2023.

\end{thebibliography}

\end{document}